\documentclass[review,sort&compress,3p]{elsarticle} 
\usepackage{amsfonts}
\usepackage{braket}
\usepackage{subcaption}
\usepackage{bm}
\usepackage[section]{placeins}
\usepackage[english]{babel}
\usepackage{amsmath, amsthm, amssymb}
\begin{document}

\title{Fast collective oscillations and clustering phenomena in an antiferromagnetic mean-field model}

\author[1,5]{Arthur Vesperini}\ead{arthur.vesperini@student.unisi.it}

\author[3]{Roberto Franzosi}\ead{roberto.franzosi@ino.it}

\author[4,2]{Stefano Ruffo}\ead{ruffo@sissa.it}

\author[4,6]{Andrea Trombettoni}\ead{andreatr@sissa.it}

\author[5]{Xavier Leoncini}\ead{xavier.leoncini@cpt.univ-mrs.fr}

\address[1]{Dipartimento di Scienze Fisiche, della Terra e dell’Ambiente (DSFTA), University of Siena, Via Roma 56, 53100 Siena, Italy}
\address[3]{QSTAR \& CNR — Istituto Nazionale di Ottica, Largo Enrico Fermi 2, I-50125 Firenze, Italy}
\address[4]{Scuola  Internazionale  Superiore  di  Studi  Avanzati  (SISSA),  I-34136  Trieste,  Italy}
\address[5]{Aix Marseille Univ, Universite de Toulon, CNRS, CPT, Marseille, France}
\address[2]{INFN Trieste and ISC-CNR Firenze }
\address[6]{CNR-IOM  DEMOCRITOS  Simulation  Center,  I-34136  Trieste,  Italy}
\date{\today}

\begin{abstract}
We study the out-of-equilibrium properties of the antiferromagnetic Hamiltonian Mean-Field model at low energy. In this regime, the Hamiltonian dynamics exhibits the presence of a stationary state where the rotators are gathered in a bicluster.
This state is not predicted by equilibrium statistical mechanics in the microcanonical ensemble. 
Performing a low kinetic energy approximation, we derive the explicit expression of the magnetization vector as a function of time. We find that the latter displays coherent oscillations, and we show numerically that the probability distribution for its phase is bimodal or quadrimodal. 
We then look at the individual rotator dynamics as a motion in an external time-dependent potential, given by the magnetization. This dynamics exhibits two distinct time scales, with the fast one associated to the oscillations of the global magnetization vector.
Performing an average over the fast oscillations, we derive an expression for the effective force acting on the individual rotator. This force is always bimodal, and determines a low frequency oscillation of the rotators.
Our approach leads to a self-consistent theory linking the time-dependence of the magnetization to the motion of the rotators, providing a heuristic explanation for the formation of the bicluster.
\end{abstract}

\maketitle

\section{Introduction}\label{sec:intro}

The Hamiltonian Mean Field (HMF) model has raised much attention in the last two decades \cite{campa_statistical_2009,antoni_clustering_1995,barre_birth_2002,barre_out--equilibrium_2002,campa_dynamics_2006,dauxois_hamiltonian_2002,dauxois_violation_2000,jeong_collective_2006,leoncini_out_2009,yamaguchi_stability_2004}. This simple toy model indeed exhibits a plethora of phenomena going beyond the scope of equilibrium statistical mechanics, as it is typically the case in long-range interacting systems.\\
As an interesting physical interpretation, the HMF model can be seen as the first Fourier mode approximation of sheet models in one-dimension; the antiferromagnetic HMF model corresponds to a charged sheets model, while the ferromagnetic HMF model corresponds to a massive sheets model \cite{antoni_clustering_1995,dawson_onedimensional_1962,eldridge_onedimensional_1962}.

The ferromagnetic HMF undergoes a second order phase transition in both the canonical and microcanonical ensembles \cite{campa_statistical_2009,vesperini_dynamics_2020}, while, to the best of our knowledge, there exists no equilibrium phase transition in the antiferromagnetic HMF. However, both the ferromagnetic and the antiferromagnetic HMF are known to present a variety of quasi-stationary states, with relaxation times diverging with the size of the system, thus entailing ergodicity breaking \cite{campa_statistical_2009,campa_dynamics_2006,antoni_clustering_1995,barre_out--equilibrium_2002,dauxois_hamiltonian_2002,leoncini_self-organized_2015}.\\
More recently, promising generalizations of the HMF model have been proposed, in which some of these interesting features can be preserved.
Notably, some of the aforementioned quasi-stationary states have been shown to be robust with respect to the addition of a (small enough) nearest-neighbours coupling to the model. Phase transition phenomena are still observed in this extended framework \cite{campa_dynamics_2006,vesperini_dynamics_2020}.  
Finite-range versions of the HMF model have also been considered \cite{turchi_emergence_2015,de_nigris_emergence_2013}, as well as extensions with higher dimensional spins \cite{dauxois_hamiltonian_2002,leoncini_self-organized_2015}, and quantum versions \cite{plestid_violent_2018}, still presenting a rich phenomenology, namely the emergence of non-trivial collective behaviours.\\

Let us introduce the Hamiltonian of the model.
We consider an assembly of $N$ planar classical rotators, endowed with a kinetic energy, subjected to an infinite-range ``antiferromagnetic" coupling. This system can also be seen as a collisionless plasma in a one-dimensional ring, with an all-to-all repulsive interaction \cite{barre_birth_2002}.
The Hamiltonian coordinates of the rotators are $\{\theta_j,p_j\}$.
The model can be defined through the Hamiltonian
\begin{equation} \begin{split}\label{eq:HMF_pot}
H & =\sum_{i=1}^N\frac{p_i^2}{2} + V(\{\theta_i\}) \text{ ,}\\
\text{with } V(\{\theta_i\}) & =\frac{1}{2N}\sum_{i,j=1}^N \cos{}(\theta_i - \theta_j) = \frac{N\bm{M}^2}{2}\, \text{,}
\end{split} \end{equation}
where $\bm{M}$ is the magnetization vector per rotator, defined as
\begin{equation}\label{eq:mag}
    \bm{M}=\frac{1}{N}\sum_{j=1}^N\begin{pmatrix} \cos{(\theta_j)}\\ \sin{(\theta_j)} \end{pmatrix}\, .
\end{equation}
The equations of motions are
\begin{equation}\label{eq:eq_of_motion}
    \dot{p}_j(t) = M_x\sin{(\theta_j)} - M_y\cos{(\theta_j)}\, .
\end{equation}
The potential of this Hamiltonian is self-consistent, a feature characteristic of mean-field models: the magnetization depends on the single rotator dynamics, which in turn depends on the former.

A homogeneous distribution of the angles of the rotators, implying a vanishing magnetization, is expected at equilibrium in both the canonical and the microcanonical ensemble. However, numerical studies have shown that a long-living coherent structure, namely a bicluster, can spontaneously form in the Hamiltonian dynamics at low energy \cite{antoni_clustering_1995,barre_birth_2002,barre_out--equilibrium_2002,dauxois_violation_2000,jeong_collective_2006}.
This quasi-stationary state consists in the gathering of an extensive quantity of rotators on two opposite angles, and is quantified by the norm $M_2$ of the vector
\begin{equation}\label{eq:mag2}
    \bm{M_2}=\frac{1}{N}\sum_{j=1}^N\begin{pmatrix} \cos{(2\theta_j)}\\ \sin{(2\theta_j)} \end{pmatrix}\, .
\end{equation}
The parameter $M_2$ varies from $0$ in the homogeneous state, to $1$ in a bicluster state with no dispersion of the rotators \cite{dauxois_violation_2000,antoni_clustering_1995,barre_birth_2002}.

Notably, bicluster states are also characterized by 
a non-zero magnetization, with $M\sim\sqrt{e}$, where $e=E/N$ is the total energy density, with $E=H(\{\theta_i(0),p_i(0)\})$. Using the kinetic definition of the temperature  $T=\braket{p^2}$, this entails an anomalous energy-temperature relation, with respect to the expected equilibrium linear relation $T=2e$ \cite{dauxois_violation_2000,antoni_clustering_1995}. \\

Remark that this phenomenon is not compatible with linear stability analysis of the Vlasov equation \cite{campa_statistical_2009,barre_birth_2002}\footnote{In Ref. \cite{campa_statistical_2009}, linear stability analysis was performed for the ferromagnetic model. Stability in the antiferromagnetic case can be retrieved by a simple change of sign.}, which predicts the homogeneous states to be stable for all energies, for a wide class of initial distributions of momenta.\\

The class of initial conditions leading to a bicluster is yet not precisely known. 
Let $\gamma_0=V_0/E$, with $V_0=V(\{\theta_i(0)\})$. Previous studies \cite{jeong_collective_2006} have shown that, at a given energy, for initially uniformly random distributions of angles and momenta (i.e. waterbag distributions, defined later in Sec.~\ref{sec:numerical_results}) the closest we are to $\gamma_0=1$, the larger is the stationary value of $M_2$. We chose to use this ratio as a control parameter for our simulations in section \ref{sec:numerical_results}. \\
Nevertheless, it is worth noting that biclusters can also arise from sinusoidal initial distributions of momenta (i.e. $p_i(0)\propto \sin{(\theta_i(0))}$), in which case the parameter $\gamma_0$ becomes irrelevant \cite{jeong_collective_2006}. In the present work, we will solely focus on waterbag initial distributions.\\

Previously, a theory has been devised to explain the formation and stabilization of a bicluster, as the equilibrium state of an averaged Hamiltonian \cite{barre_birth_2002}, derived by using a variational method inspired by Ref. \cite{whitham_linear_2011}. The authors of Ref. \cite{barre_birth_2002}, separating fast and slow variables in the Lagrangian, notably predicted the occurrence of two collective high frequencies $\omega_\pm$, and gave accurate quantitative results.\\

In the following, we propose a new approach, to get a better understanding of the dynamical mechanism at the base of the bicluster formation and stabilization. 
We derive the same high frequencies $\omega_\pm$ in section \ref{sec:mag_dyn}, by directly studying the dynamics of the magnetization vector, which is the driving force of the system (see Eq.~\eqref{eq:eq_of_motion}).
This allows us, in section \ref{sec:time_averaging}, to rewrite the equations of motions in a non-autonomous form, and thereby perform an averaging over the fast variables in a very simple fashion. An expression for the effective force is found, with associated low frequency $\omega_0$, and its dependence to initial conditions is discussed.\\
Section \ref{sec:numerical_results} exposes our numerical results, showing excellent agreement with the theory.\\
In section \ref{sec:discussion}, we discuss our results and develop a heuristic argument to explain the birth and stabilization of the bicluster states. We conclude by mentioning possible analogies with other models, and proposing further developments. 

\section{Dynamics of the total magnetization}\label{sec:mag_dyn}

We are interested in deriving a dynamical equation for the macroscopic quantity $\bm{M}$. From Eq.~\eqref{eq:mag}, we get
\begin{equation}
    \frac{d^2}{dt^2}\bm{M}(t) = \frac{1}{N}\sum_{j=1}^N\begin{pmatrix}  -\dot{p}_j\sin{(\theta_j)} - p_j^2\cos{(\theta_j)} \\  \dot{p}_j\cos{(\theta_j)}  -p_j^2\sin{(\theta_j)} \end{pmatrix}\, .
\end{equation}
We identify in this expression the correlator $\braket{p^2\cos{(\theta)}}\sim o(T)$, that we can neglect in the small temperature regime. We are left with
\begin{equation}
    \frac{d^2}{dt^2}\bm{M}(t)
\approx\frac{1}{N}\sum_{j=1}^N \begin{pmatrix} -\dot{p}_j\sin{(\theta_j)} \\ \dot{p}_j\cos{(\theta_j)} \end{pmatrix}\, .
\end{equation}
Then, inserting the equations of motion \eqref{eq:eq_of_motion}, we obtain the eigenproblem
\begin{equation}\label{eq:eipbM}
        \frac{d^2}{dt^2}\begin{pmatrix} M_x(t) \\ M_y(t) \end{pmatrix} \approx \begin{pmatrix} -\frac{1-M^{(2)}_x}{2} & \frac{M^{(2)}_y}{2} \\ \frac{M^{(2)}_y}{2} & -\frac{1+M^{(2)}_x}{2} \end{pmatrix}\begin{pmatrix} M_x(t) \\ M_y(t) \end{pmatrix}\, .
\end{equation}
The eigenvalues and corresponding eigenvectors result
\begin{gather}
        -\omega_\pm^2 = -\frac{1\pm M_2}{2}\\
        \bm{M_-}=\begin{pmatrix} \cos{(\phi_2/2)} \\ \sin{(\phi_2/2)} \end{pmatrix} \text{,  }\; \bm{M_+}=\begin{pmatrix} -\sin{(\phi_2/2)} \\ \cos{(\phi_2/2)} \end{pmatrix} \text{ , }\label{eq:M_eivectors}
\end{gather}
where $\phi_2$ is defined as the phase of $\bm{M_2}$. We hence expect the system to globally rotate with $\phi_2/2$, which already stresses the importance of $\bm{M_2}$ in the characterization of the dynamics. \\
Let us emphasize the consistency of this result with that one of Ref. \cite{barre_birth_2002}, in which the modes $\omega_\pm$ were found to be the eigenvalues of the fast Lagrangian, and where $\phi_2/2$ was already recognized as the system's center of mass. These frequencies, arising from nonlinear mode interaction, can be seen as a splitting of the single normal mode $\omega=1/\sqrt{2}$, present in the homogeneous state \cite{jeong_collective_2006}. This normal mode can also be found by linear analysis of the Vlasov equation \cite{barre_birth_2002}.\\

Assuming $\bm{M_2}$ constant, after a global rotation of $-\phi_2/2$, we get
\begin{equation}\label{eq:M_gen_solution}
\bm{M} = \begin{pmatrix}
M_-\cos{(\omega_-t+\phi_-)} \\M_+\cos{(\omega_+t+\phi_+)}\end{pmatrix}\, .
\end{equation}

It is worth remarking that, in the ferromagnetic case, the eigenvalues read $\lambda_\pm\approx \frac{1\pm M_2}{2}$, and under the same low energy hypothesis, $\bm{M}$ will rather converge to a constant, following a slow drift motion \cite{antoni_clustering_1995}.\\

\begin{figure}[ht]
    \centering
     \begin{subfigure}[t]{.45\textwidth}
        \centering 
       \includegraphics[height=.5\textwidth]{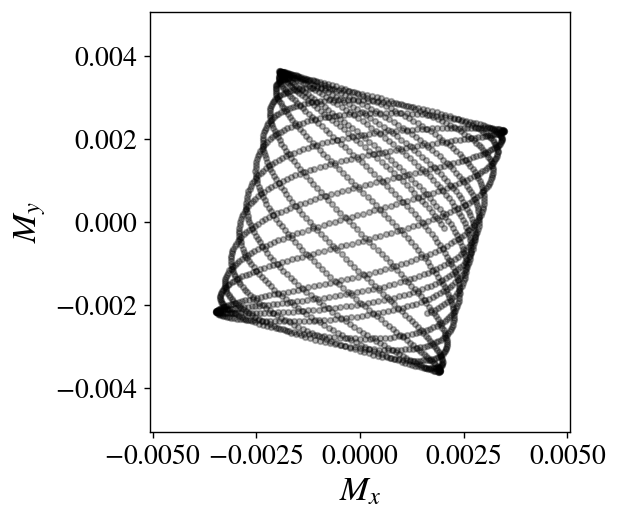}\caption{}\label{fig:MxMy-portrait}
     \end{subfigure}
     \begin{subfigure}[t]{.45\textwidth}
         \centering
        \hspace*{-4cm}\includegraphics[height=.5\textwidth]{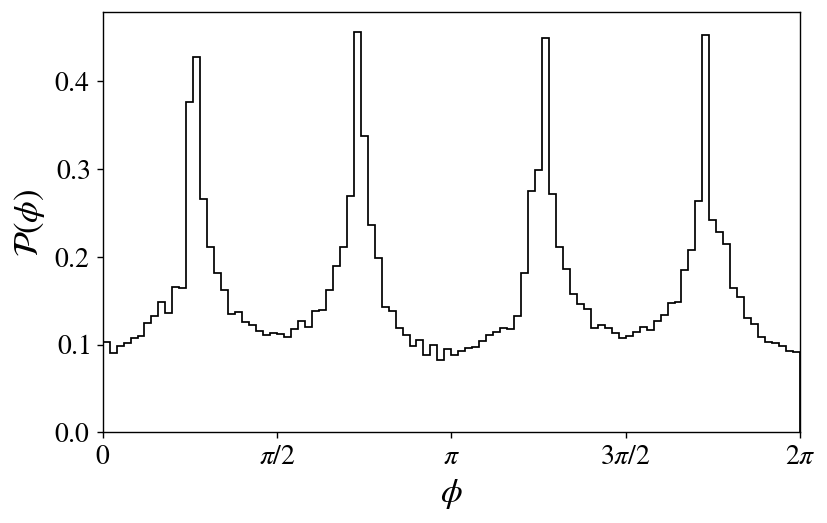}\caption{}
         \label{fig:M-hist}
     \end{subfigure}
     \vskip\baselineskip
     \centering
     \begin{subfigure}[t]{.45\textwidth}
         \centering
         \includegraphics[height=.5\textwidth]{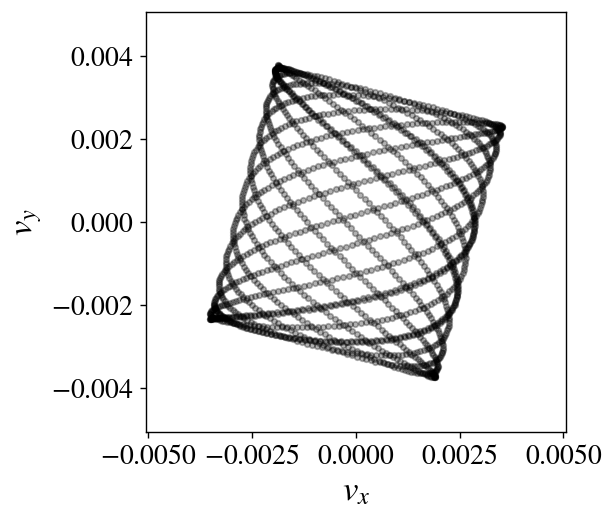}
            \caption{} \label{fig:vxvy-portrait}
     \end{subfigure} 
     \begin{subfigure}[t]{.45\textwidth}
         \centering
         \hspace*{-4cm}\includegraphics[height=.5\textwidth]{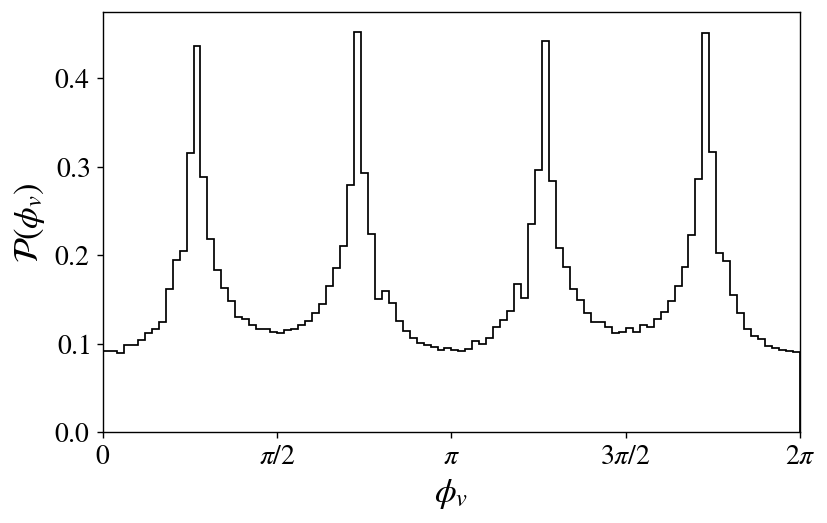}
         \caption{}
         \label{fig:v-hist}
     \end{subfigure}
     \caption{The dynamics and histogram of $\phi$, the phase of $\bm{M}$, measured in a simulation with $\gamma_0=1$, $M_2=0.52$ (see Sec.~\ref{sec:numerical_results}, Fig.~\ref{fig:traj_withM_bic1}).
     We define a vector $\bm{v}$ according to Eq.~\eqref{eq:M_gen_solution}. The frequencies $\omega_\pm$ are drawn from measurements of $M_2$, the amplitudes $v_{-,+}$ respectively defined through $\max(M_{x,y})$ (after $\bm{M}$ was rotated of $-\phi_2/2$), and we finally rotate $\bm{v}$ of $\phi_2/2$, as suggested by Eq.~\eqref{eq:M_eivectors}.
    We show in Cartesian coordinates the dynamic, measured over the time interval $t\in[10000,10100]$, of $\bm{M}$ (resp. $\bm{v}$) in Fig.~\ref{fig:MxMy-portrait} (resp. \subref{fig:vxvy-portrait}). 
    The corresponding distributions $\mathcal{P}(\phi)$ (resp. $\mathcal{P}(\phi_v)$) are reported in Fig.~\ref{fig:M-hist} (resp. \subref{fig:v-hist}). Histograms are derived from a sample of values retrieved in the time interval $t\in[10000,11000]$. We used the time step $\Delta t=0.05$. }
    \label{fig:M-v_compare}
\end{figure}

Fig.~\ref{fig:M-v_compare} shows the behaviour of $\bm{M}$ measured by numerical integration of the full equations of motion \eqref{eq:eq_of_motion}, and the one of a vector $\bm{v}$ defined according to Eq.~\eqref{eq:M_gen_solution}. Namely, $\bm{v}=\big(v_-\cos{(\omega_-t)},v_+\cos{(\omega_+t)}\big)$, where $\omega_\pm$ are computed from the average value of $M_2$ and $v_\pm$ are measured by taking the maximum value of $M_{x,y}$, after a rotation of $-\phi_2/2$. Also, we present later, in Fig.~\ref{fig:Mxy_spectrum} the frequency spectrum of the components $M_x$, $M_y$, derived by a fast Fourier transform, also performed after a rotation of $-\phi_2/2$. Better agreement is found for well-formed biclusters ($M_2\gtrsim0.1$), as we will discuss in section \ref{sec:numerical_results}. 
A few other examples are displayed in \ref{apx:M-v-compare_appendix}.

The parametric curves defined by Eq.~\eqref{eq:M_gen_solution} are named Lissajous curves. Such curves are bounded in the $(M_x,M_y)$-plane by a rectangle of sides $M_-$ and $M_+$, and are known to densely fill its area, provided that the ratio $\omega_-/\omega_+$ is irrational, condition that is almost always fulfilled. 
The norm $M$ evidently possesses four maxima, each located at a fixed angular position. 
One can see from Eq.~\eqref{eq:M_gen_solution} that $\dot{\bm{M}}$ approaches $0$ as $\bm{M}$ approaches $(\pm M_-,\pm M_+)$ (when $M$ is maximal), making these regions favoured in terms of the amount of time spent there by the system, as illustrated in Fig.~\ref{fig:M-hist}.\\
Note that, when $M_2\approx0$, the curve is an ellipse, hence $\bm{M}$ exhibits two maxima. This is also the case when $M_-\ll M_+$ (or $M_-\ll M_+$).
The probability density function for $\bm{M}$ is hence bimodal (at least during a first transient phase) or quadrimodal.\\

This simple derivation already provides us with a heuristic explanation for the occurrence of a bimodal distribution of the rotators in the antiferromagnetic HMF model. 
Indeed, as we will show below, if the rotators are slow enough with respect to $\omega_\pm$, they effectively experience a bimodal potential.\\ 
It is fairly obvious from Fig.~\ref{fig:M-v_compare} that, as this regime persists for very long times, it brings on a breaking of ergodicity. Indeed, the accessible state space is bounded by the Lissajous curve, entailing a probability density $\mathcal{P}(\phi,M)$ anomalous with respect to the expected one from equilibrium statistics.
In particular, while the time average of $\bm{M}$ is null, the one of $M$ is not. These coherent oscillations hence allow for a non-vanishing (extensive) average potential energy. 

\section{Time scale separation}\label{sec:time_averaging}
We found above an explicit time dependence for the bare potential. By doing this, we also decoupled it from the generalized coordinates $\{\theta_j\}$. This allows us to fully take advantage of the mean-field nature of the model, hence to actually consider single rotators as uncoupled pendula, evolving under the action of an external potential driven by the oscillating ``magnetic field" $\bm{M}$.

We first insert Eq.~\eqref{eq:M_gen_solution} in Eq.\eqref{eq:eq_of_motion}, thus 
\begin{equation}\label{eq:eq_of_motion_time_dep}
    \dot{p}_j(t) = \epsilon a_-\cos{(\omega_- t + \phi_-)}\sin{(\theta_j)} - \epsilon a_+\cos{(\omega_+ t + \phi_+)}\cos{(\theta_j)} \, \text{,}
\end{equation}
with $\epsilon a_\pm = M_\pm$, so that we have $a_+^2 + a_-^2=1$, and $\epsilon = \sqrt{M_+^2 + M_-^2}\sim\sqrt{e}$. At low energy, two distinct time scales arise from this expression: a large one, associated to $\sqrt{\epsilon}$, and a small one, associated to the high frequencies $\omega_\pm \sim 1$. 

We are now able to perform a simple approximation, related to the \textit{ponderomotive effect}, well-known in the area of plasma physics \cite{schmidt_ii_1979}. As the one employed in Ref. \cite{barre_birth_2002}, it relies on the clear separation of time scales between fast and slow variables, and is somehow analogous to the method first proposed by Landau and Lifshitz to solve systems exhibiting two distinct time scales \cite{landau_chapter_1976}. The prototypical example of such systems is the Kapitza pendulum \cite{kapitza_collected_1965}.\\
Let us decompose the variables in a fast and a slow component. We set the magnitude of the fast component to be $o(\epsilon)$, and introduce a ``slow time" $\tau=\epsilon t$, associated to the slow oscillations, insuring $\braket{\dot{p}_j^2} \sim \epsilon^2$,
\begin{equation}
    \theta_j (t) = \theta^0_j (\tau,t) + \epsilon f_j(t) \, .
\end{equation}
The single rotator dynamics thus presents a fast motion of small amplitude, superimposed with a slow motion of large amplitude.

Expanding Eq.~\eqref{eq:eq_of_motion_time_dep} up to first order in $\epsilon f_j$, we obtain
\begin{equation}\label{eq:eq_of_motion_expanded}
\begin{split}
    \epsilon^2\frac{d^2}{d\tau^2}\theta_j^0(\tau,t) + \epsilon\frac{d^2}{dt^2}f_j(t)
    = &\epsilon \Big(a_-\cos{(\omega_- t + \phi_-)}\sin{(\theta^0_j(\tau))} - a_+\cos{(\omega_+ t + \phi_+)}\cos{(\theta^0_j(\tau))} \Big)\\
    & + \epsilon^2 f_j(t) \Big(a_-\cos{(\omega_- t + \phi_-)}\cos{(\theta^0_j(\tau))} + a_+\cos{(\omega_+ t + \phi_+)}\sin{(\theta^0_j(\tau))} \Big)  \, .
\end{split}
\end{equation}
By identifying terms order by order, we get the following expression for the fast variables
\begin{equation}
    \frac{d^2}{dt^2}f_j(t)= a_-\cos{(\omega_- t + \phi_-)}\sin{(\theta^0_j(\tau))} - a_+\cos{(\omega_+ t + \phi_+)}\cos{(\theta^0_j(\tau))} \text{ ,}
\end{equation}
which we can straightforwardly integrate, since $\theta^0_j(\tau)$ is considered constant on the time scale of $f_j(t)$. It results
\begin{equation}\label{eq:fast_variable}
    f_j(t)= -\frac{a_-}{\omega_-^2}\cos{(\omega_- t + \phi_-)}\sin{(\theta^0_j(\tau))} +\frac{a_+}{\omega_+^2}\cos{(\omega_+ t + \phi_+)}\cos{(\theta^0_j(\tau))} \, .
\end{equation}
Then, by substituting this expression for $f_j(t)$ in Eq.~\eqref{eq:eq_of_motion_expanded}, we obtain after some manipulations (for convenience, we dropped the time dependence and the constant phases $\phi_\pm$)
\begin{equation}
\begin{split}
        \frac{d^2}{dt^2}\theta^0_j=&\frac{1}{4}\Bigg[\frac{M_+^2}{\omega_+^2}\Big(1+\cos{(2\omega_+t)}\Big)-\frac{M_-^2}{\omega_-^2}\Big(1+\cos{(2\omega_-t)}\Big)\Bigg]\sin{(2\theta^0_j)}\\
        &+\frac{1}{4}\Bigg[\frac{M_+M_-}{\omega_+^2}\Big(1+\cos{(2\theta^0_j)}\Big)-\frac{M_+M_-}{\omega_-^2}\Big(1-\cos{(2\theta^0_j)}\Big)\Bigg]\Big(\cos{((\omega_-+\omega_+) t)}+\cos{((\omega_--\omega_+)t )}\Big) \,.
\end{split}
\end{equation}
If $M_2$ is of the order of $\epsilon$ then $(\omega_+-\omega_-)$ is of the order of $\epsilon$, a low frequency that cannot be neglected by averaging over the fast oscillations. Then, our computation holds when $M_2\gg\epsilon$, and a priori does not account for the beginning of the transient.
By averaging over the fast oscillations, we get the expression for the slow variables
\begin{equation}\label{eq:slow_var_general}
    \frac{d^2}{dt^2}\theta^0_j \approx \frac{1}{4}\Big( \frac{M_+^2}{\omega_+^2}-\frac{M_-^2}{\omega_-^2}  \Big)\sin{(2\theta^0_j)}\, .
\end{equation}

Assuming that the prefactor is negative, we can consider a rotator in the bottom of one potential well, located at $\theta^0_j\approx k\pi$, with $k\in\mathbb{Z}$, so $\sin{(2\theta^0_j)}\approx 2\theta^0_j - 2k\pi$. We then have
\begin{gather}
    \theta^0_j(t)\approx k\pi + A_j\cos{(\omega_0 t + \phi_j)} \text{, with}\label{eq:slow_var_in_well}\\
    \omega_0 = \frac{1}{\sqrt{2}}\sqrt{ \frac{M_-^2}{\omega_-^2}-\frac{M_+^2}{\omega_+^2} } \, .\label{eq:omega0}
\end{gather}

$\omega_0$ is of the order of $M$, namely the square of the natural frequency. This emphasizes that the effective force emerges from the non-linearity, linked to the self-consistency of the magnetization.

The attractive or repulsive nature of this bimodal effective interaction, is related to the sign of the prefactor in Eq.~\eqref{eq:slow_var_general}, namely
\begin{equation}\label{eq:attrac_int_cond}
\Delta - M_2 < 0 \text{ ,}   
\end{equation} with $\Delta = \frac{M_+^2-M_-^2}{M_+^2+M_-^2}$. The initial value of the latter is unrelated to the control parameter $\gamma_0$. But a bicluster should not be stable unless the condition \eqref{eq:attrac_int_cond} is satisfied. Yet we know that the biclusters are stable, provided that $\gamma_0$ is large enough. 
Moreover, the effective force is self-consistent, in the sense that its strength is proportional to $M_2$, which is itself governed by the former.
Thus, we are brought to assume that $M_+$, $M_-$ and $M_2$ are evolving during a transient phase in an interdependent fashion, following a dynamics somehow determined by $\gamma_0$.

This result provides a dynamical explanation for the stabilization of biclusters over very long times.

\section{Numerical results}\label{sec:numerical_results}
Our simulations were performed at energies ranging from $10^{-5}$ to $10^{-4}$, with $N=1000$. The equations of motions have been integrated using a fourth-order sympleptic scheme \cite{mclachlan_accuracy_1992}. For most of the figures, we used a time step $\Delta t=0.05$, which gives a conservation of the energy up to $\Delta e\sim10^{-12}$. On the contrary, to produce Fig.~\ref{fig:om0_th_vs_exp}, we used a more efficient time step $\Delta t=0.5$, yielding $\Delta e\sim10^{-6}$. For the purpose of measuring a low frequency, with an efficient integrating scheme and at these ranges of energy, such a time step remains of an acceptable precision.\\
We initially set a water-bag distribution, picking the positions and momenta uniformly at random in a domain $[-\pi,\pi]\times[-p_0,p_0]$.
We used the prescription of Ref. \cite{jeong_collective_2006} to use $\gamma_0$ as a control parameter. To do this, we first find, by iterating multiple times, a distribution of positions giving a potential energy in the desired range. Then we choose $p_0$ to set $T=\braket{\dot{\theta_j}^2}$ accordingly, and globally shift the momenta to set the constant of motion $\braket{p_j} = 0$.\\
Measurements are performed at time $t>10000$ to insure that the system has passed the transient and reached a steady state. 
\begin{figure}[ht!]
    \centering
     \begin{subfigure}[t]{.45\textwidth}
         \centering
         \includegraphics[width=\textwidth]{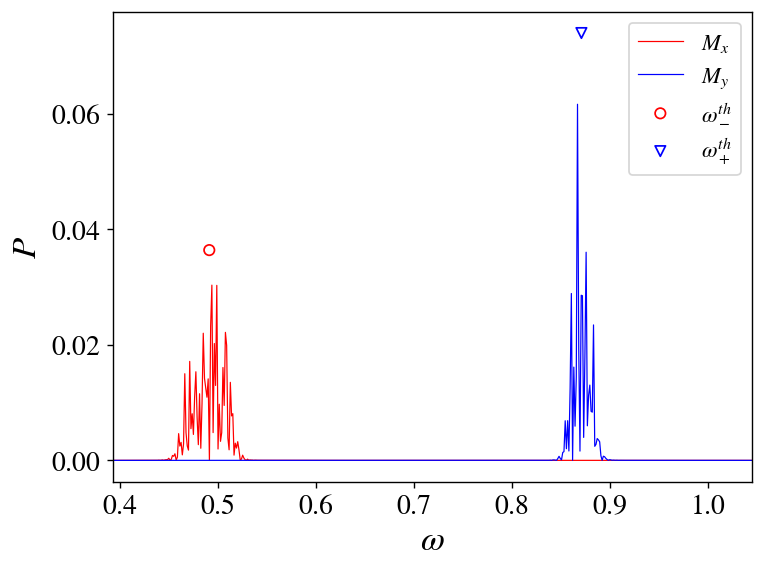}
         \caption{Frequency spectrum of $M_x$ and $M_y$.}
         \label{fig:Mxy_spectrum}
     \end{subfigure}
          \hfil
     \begin{subfigure}[t]{.45\textwidth}
         \centering
         \includegraphics[width=\textwidth]{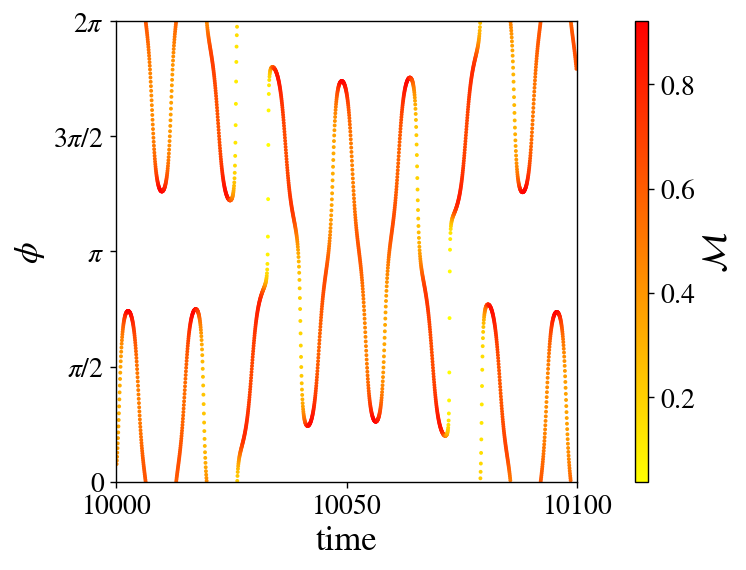}
            \caption{Short time phase trajectory of $\bm{M}$.}  \label{fig:short_t_phi}
     \end{subfigure} 
     \caption{Power spectrum and detail of the trajectory of $\bm{M}$, with $\gamma_0=1$, $M2=0.52$, $\Delta=-0.09$ (see Fig.~\ref{fig:traj_withM_bic1}).}
    \label{fig:M_features_bic}
\end{figure}

In the literature \cite{barre_birth_2002,dauxois_violation_2000,antoni_clustering_1995}, as well as in our own simulations, the parameter $M_2$ has never been reported to exceed $0.8$.\\

In Fig.~\ref{fig:M_features_bic} is shown an example of the short term dynamics of $\bm{M}$, along with the corresponding Fourier spectra of its components, performed after a global rotation of $-\phi_2/2$. Here, the agreement of experimental data with Eq.~\eqref{eq:M_gen_solution} is excellent.\\
For small $M_2$, the agreement of $\bm{M}$ with Eq.~\eqref{eq:M_gen_solution} is not as good. Though collective oscillations still occur, the envelopes $M_\pm$ fluctuate, and the trajectories of the magnetization lose their regularity.\\
However, we observed the fast collective oscillations to be present from the beginning, regardless of the later formation of a bicluster (hence of the value of $\gamma_0$), and before the system has reached a stationary state. \\
The average value of $\mathcal{M}=M/\sqrt{2e}$ is related to $\gamma_0$: a high initial value leads to an accordingly high average of $\braket{\mathcal{M}}$. \\

\begin{figure}[ht!]
    \centering
     \begin{subfigure}[t]{.45\textwidth}
        \centering
         \caption{$\gamma_0=1$, $M2=0.52$, $\Delta=-0.09$}
         \label{fig:traj_withM_bic1}
         \includegraphics[width=\textwidth]{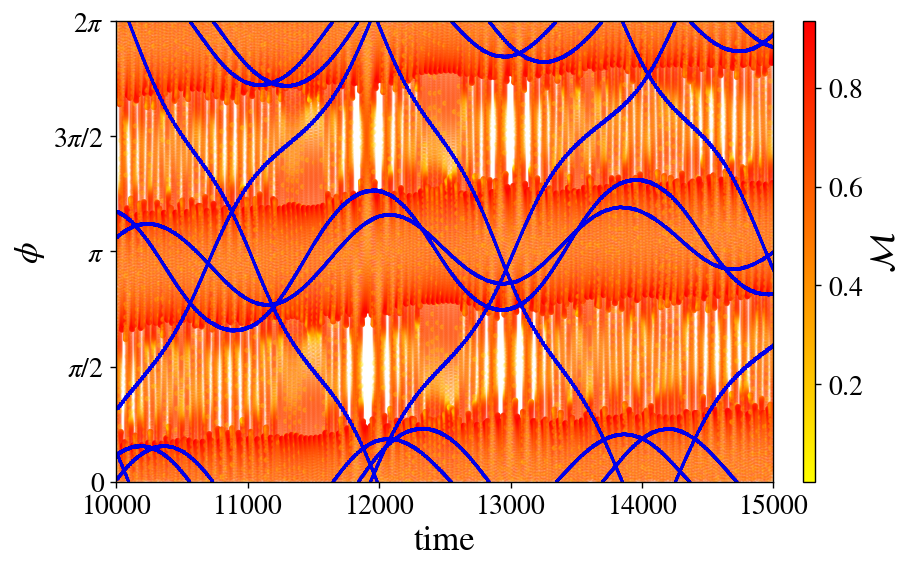}
     \end{subfigure}
     \hfil
     \begin{subfigure}[t]{.45\textwidth}
         \centering
            \caption{$\gamma_0=1$, $M_2=0.51$, $\Delta=-0.98$}     \label{fig:traj_withM_2_bic2}
         \includegraphics[width=\textwidth]{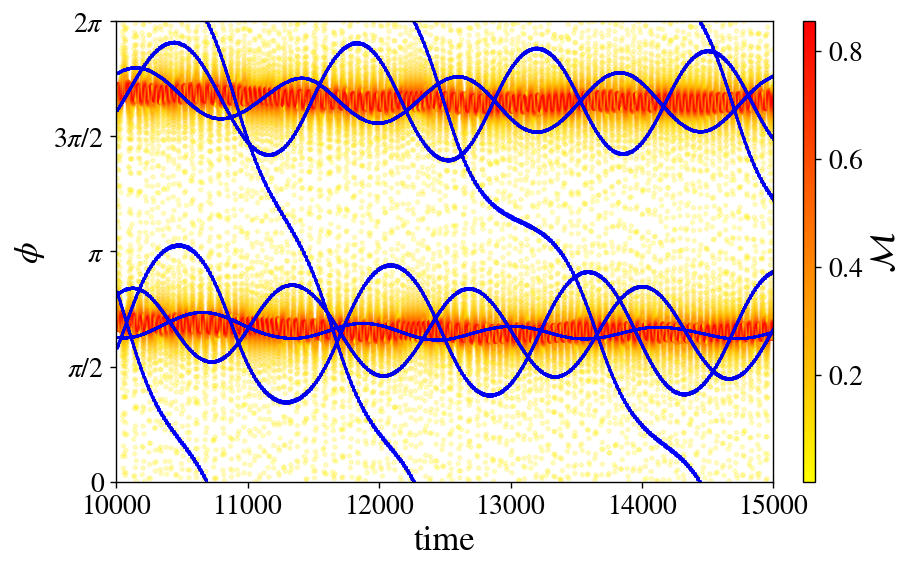}
     \end{subfigure} 
     \vskip\baselineskip
    \begin{subfigure}[t]{.45\textwidth}
         \centering
            \caption{$\gamma_0=0.46$, $M_2=0.06$, $\Delta=-0.94$}   \label{fig:traj_withM_dis1}
         \includegraphics[width=\textwidth]{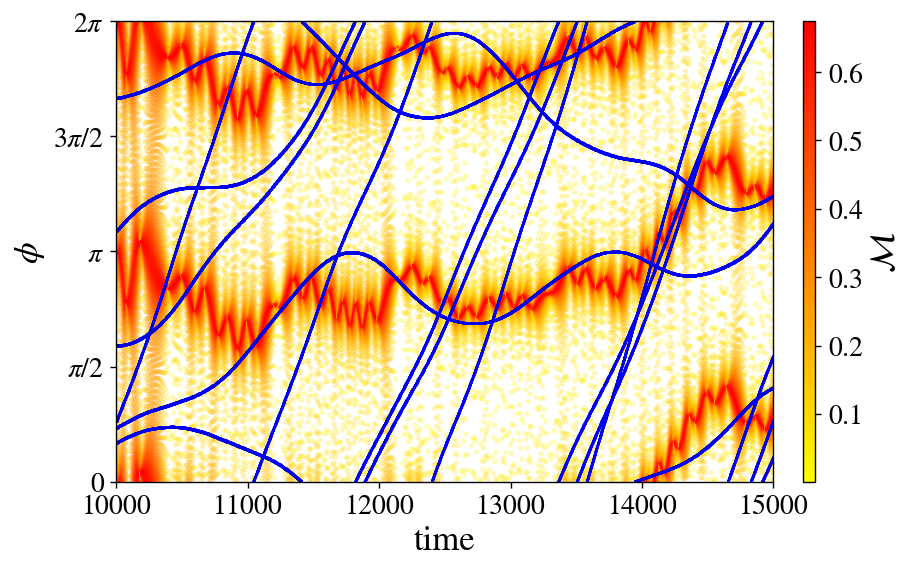}
     \end{subfigure} 
     \hfil
    \begin{subfigure}[t]{.45\textwidth}
         \centering            \caption{$\gamma_0=0.05$, $M_2=0.03$, $\Delta=-0.11$}  \label{fig:traj_withM_dis2}
         \includegraphics[width=\textwidth]{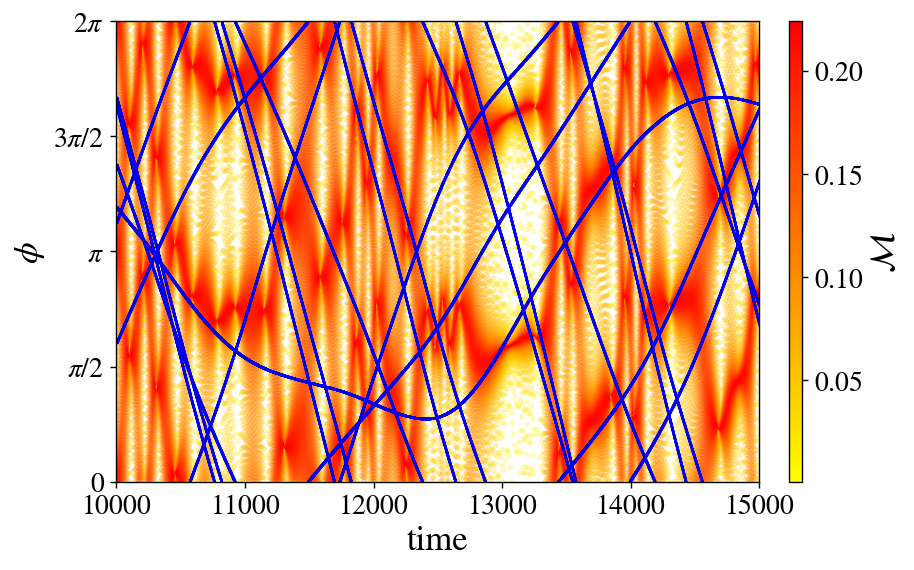}
     \end{subfigure}  
     \caption{Some rotators trajectories are shown in blue, along with the magnetization in a red gradient; the color gradient indicates the rescaled norm $\mathcal{M}=  M/\sqrt{2e}$. Energy was set to $e\sim10^{-5}$.}
    \label{fig:traj_withM}
\end{figure}
Fig.~\ref{fig:traj_withM} shows the general dynamics of the system, at different values of the parameters. 
Here, the existence of two distinct time scales is manifest: the one associated to the fast oscillation of $\bm{M}$, is visibly much smaller than the one associated to the long-term behaviour of the single rotator dynamics. In this view, it is evident that the dynamics associated to the slow variables is similar to one of a rotator in a bimodal potential.\\
Indeed, we can clearly see two angular regions ``favoured" by $\bm{M}$ in terms of the time spent as well as in magnitude. These are the locations of the two clusters, following as expected the same slow linear drift as $\phi_2/2$. 
Around these regions some trapped rotators (below the separatrix) slowly oscillate, while some untrapped ones (above the separatrix) are evolving in an almost ballistic fashion.

Note that well-formed biclusters seem to occur regardless of the value of $\Delta$. Indeed, we were not able to find a clear relation of the stationary value of $\Delta$ neither with $\gamma_0$ nor with the stationary value of $M_2$. \\
Although we have found that the effective force Eq.~\eqref{eq:slow_var_general} can become very slightly repulsive when $\gamma_0\approx0$, it ends up attractive in the vast majority of cases. Also, $\Delta$ and $M_2$ evolve, at a slow time scale with respect with $\omega_\pm$, towards values satisfying Eq.~\eqref{eq:attrac_int_cond}. \\

\begin{figure}[ht!]
\centering
\begin{minipage}[t]{.45\textwidth}
  \centering
     \includegraphics[width=\textwidth]{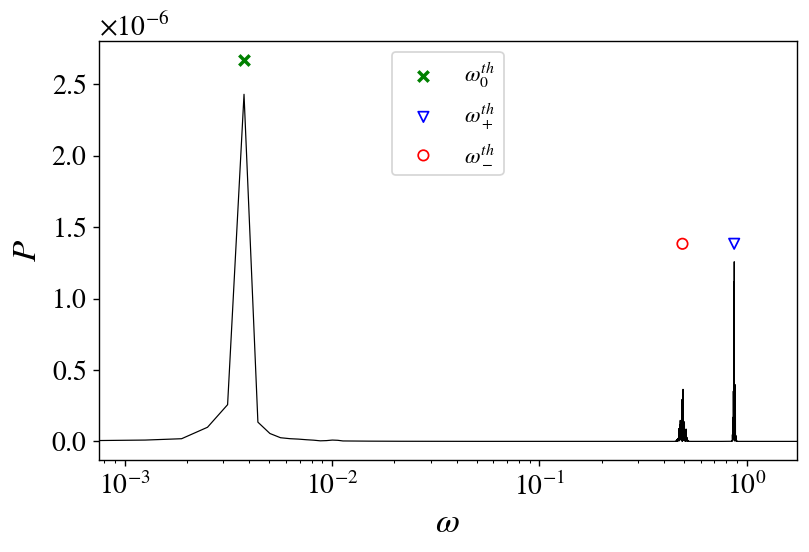}
    \caption{Power spectrum of a single trapped rotator from figure \ref{fig:traj_withM_bic1}. }
    \label{fig:rotator_spectrum}
\end{minipage}
\hfil
\begin{minipage}[t]{.45\textwidth}
  \centering
    \includegraphics[width=\textwidth]{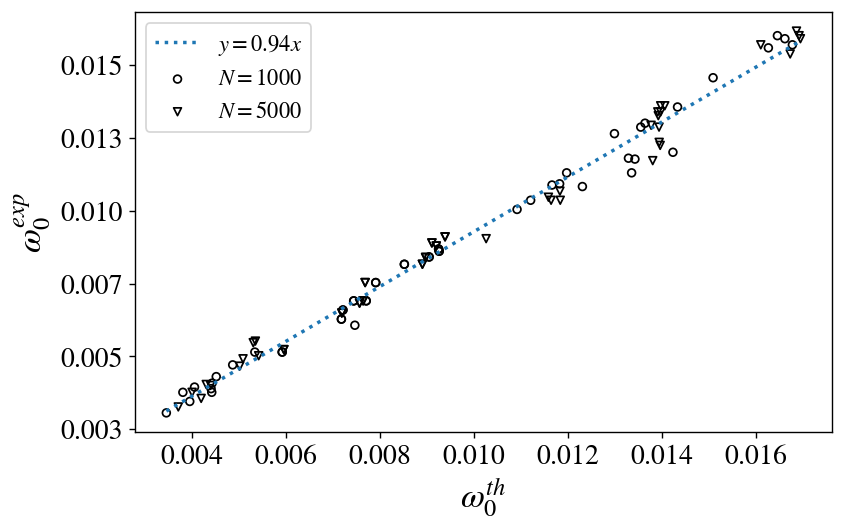}
    \caption{Comparison of $\omega_0$ theoretical and experimental. We performed Fourier transforms on small random subsets of trapped rotators, at energies ranging from $10^{-5}$ to $10^{-4}$.}
    \label{fig:om0_th_vs_exp}
\end{minipage}
\end{figure}
To investigate the spectral properties of the rotator trajectories, we focused on initial conditions leading to sufficiently well-formed biclusters, i.e. $M_2\gtrsim0.2$ ($\gamma_0>0.7$), and performed the global rotation of $-\phi_2/2$ to follow the center of mass.

Fig.~\ref{fig:rotator_spectrum} shows an example of a single rotator Fourier spectrum, trapped in a potential well and oscillating with a small amplitude. The slow mode $\omega_0$ is not present in the spectra of the untrapped rotators, or is very weak and with a higher discrepancy with Eq.~\eqref{eq:omega0}.

In the considered regime, the low frequency observed in simulations agrees with our theoretical value, up to a multiplicative factor of order $1$, namely $\omega_0^{exp}\approx0.94\omega_0^{th}$, as shown in Fig.~\ref{fig:om0_th_vs_exp}.

\section{Conclusions and perspectives}\label{sec:discussion}
In the light shed by these results, let us synthesize and propose a scenario accounting for the formation of biclusters in the antiferromagnetic HMF model, from a waterbag initial distribution.\\

At low energy, an initial state of small but non-vanishing magnetization generates a collective oscillatory regime. This is due to the self-consistency of $\bm{M}$, which repels all of the rotators, entailing its own motion towards the opposite angle, in a periodic fashion. The time scale associated to this collective motion is much smaller than the typical time scale of the individual rotators. 
We observe a cyclic high frequency transfer of energy between kinetic and potential, and the system periodically returns close to its initial high value of the $\gamma=M^2/2E$ ratio. This entails the non-vanishing $\braket{M}$.  
If, initially, the temperature is small with respect to the potential energy, the magnetization vector $\bm{M}$ follows a Lissajous-type regular curve parametrized by $\omega_\pm=\sqrt{\frac{1-M_2}{2}}$ and $M_\pm$, as described in Eq.~\eqref{eq:M_gen_solution}. The phase of $\bm{M}$ is rapidly oscillating between two or four symmetric angles, and we thus have $\braket{\bm{M}}=\bm{0}$ and, near one of the maxima, rotators are subjected to repulsive and attractive forces, alternatively.\\
In the very beginning, as $M_2\approx0$, $\omega_-\approx\omega_+\approx1/\sqrt{2}$, and $\bm{M}$ follows an almost elliptic trajectory, and thus exhibits two maxima in magnitude at two opposite angular positions. 
The variables $\omega_\pm$, $M_\pm$, are evolving concomitantly with $M_2$, at a slow rate.
As $M_2$ increases, the unique frequency of $\bm{M}$ split into two, and the two maxima (generally) split into four. \\
When the difference between the two frequencies becomes large enough, a bimodal effective force can be derived, accounting for the stabilization of the bicluster. 

The nature of this effective force is determined by Eq.~\eqref{eq:attrac_int_cond}. A full understanding of the conditions leading to a stable bicluster would thus involve a thorough study of the transient dynamics of the slow macroscopic variables $\bm{M_2}$, $M_+$ and $M_-$. \\
It would also require to explain how other types of initial distributions (in particular, initial sinusoidal distributions of momenta, with vanishing initial magnetization) relate to the processes described above.

The study of the dynamics of this simple mean-field model provides valuable insights into the mechanisms leading to ergodicity breaking in long-range interacting systems.\\
We have stressed the importance of the self-consistency of the potential, giving rise to nonlinear effects, solvable through multiscale analysis. 
This self-consistency is characteristic of mean-field models; an interesting development would hence be to look for the presence of biclusters and collective oscillations in modified versions of the antiferromagnetic HMF, weakening this self-consistency. 
This emergent behaviour has been shown to be preserved in presence of a nearest-neighbour ferromagnetic or antiferromagnetic perturbative interaction \cite{vesperini_dynamics_2020}; the phenomenon is hence not specific of pure mean-field models.\\
In recent studies, it has been noticed that the HMF model presents strong similarities with systems of cold atoms in optical cavities \cite{schutz_dissipation-assisted_2016, schutz_prethermalization_2014}. Such systems can be considered as almost isolated, thus opening the possibility of performing a ``real-life experiment" showing the non-trivial ordered phases discussed in this paper.

{\bf Acknowledgments}
R.F. acknowledges support by the QuantERA ERA-NET Co-fund 731473 (Project Q-CLOCKS).
S.R. is financially supported by the MISTI Global Seed Funds MIT-FVG Collaboration Grant ``NV centers for the test of the Quantum Jarzynski Equality (NVQJE)”, and the MIUR-PRIN2017 project ``Coarse-grained description for non-equilibrium systems and transport phenomena (CO-NEST)” No. 201798CZL.

\appendix
\section{Examples of magnetization dynamics}\label{apx:M-v-compare_appendix}
 
Below are shown the dynamics of $\bm{M}$ in Cartesian coordinates, along with the corresponding distributions $\mathcal{P}(\phi)$, from different simulations. The histograms are derived from samples of values retrieved in the time interval $t\in[10000,11000]$, while the dynamics are bounded by the time interval $t\in[10000,10100]$. We used the time step $\Delta t=0.05$. \\
The upper figures display the dynamics directly retrieved from simulations, while the lower ones shows the same views of a vector $\bm{v}$ defined using Eq.~\eqref{eq:M_gen_solution}. The frequencies $\omega_\pm$ are drawn from measurements of $M_2$, $v_\pm=\max(M_{x,y})$ (with $\bm{M}$ rotated of $-\phi_2/2$), and we finally rotate $\bm{v}$ of $\phi_2/2$, as suggested by Eq.~\eqref{eq:M_eivectors}.

Visibly, the discrepancy between the real dynamics and our analytical formula is higher for less well-formed biclusters. This is due to the fact that, as mentioned is section \ref{sec:numerical_results}, in this regime, the amplitudes $M_\pm$ are fluctuating, whereas our parameters $v_\pm$ are constant.

\begin{figure}[ht]
    \centering
     \begin{subfigure}[t]{.45\textwidth}
         \centering \includegraphics[height=.5\textwidth]{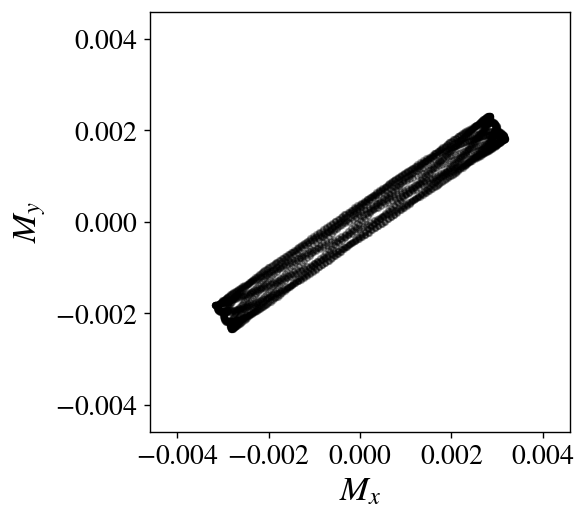}
         \caption{}
     \end{subfigure}
     \begin{subfigure}[t]{.45\textwidth}
         \centering \hspace*{-4cm}\includegraphics[height=.5\textwidth]{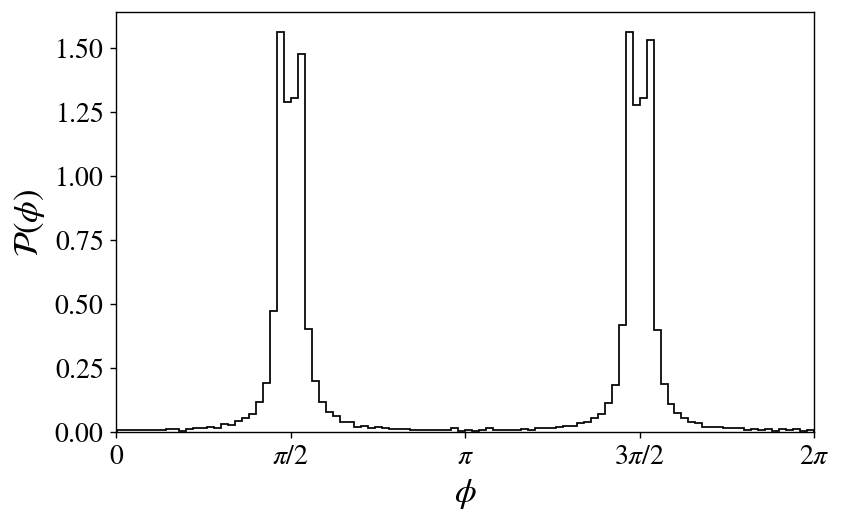}
         \caption{}
     \end{subfigure}
     \vskip\baselineskip
     \begin{subfigure}[t]{.45\textwidth}
         \centering \includegraphics[height=.5\textwidth]{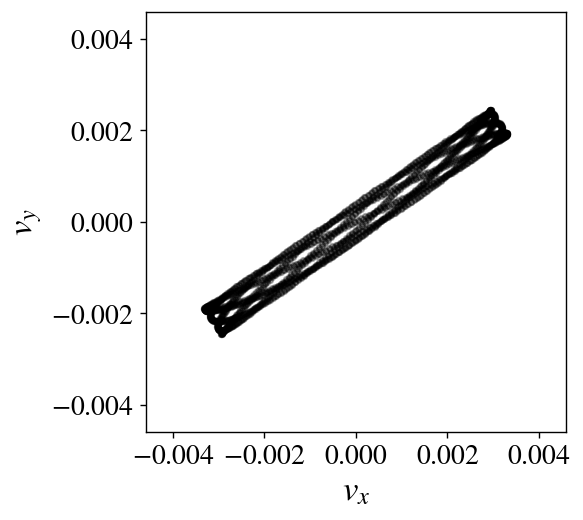}
            \caption{}
     \end{subfigure} 
     \begin{subfigure}[t]{.45\textwidth}
         \centering \hspace*{-4cm}\includegraphics[height=.5\textwidth]{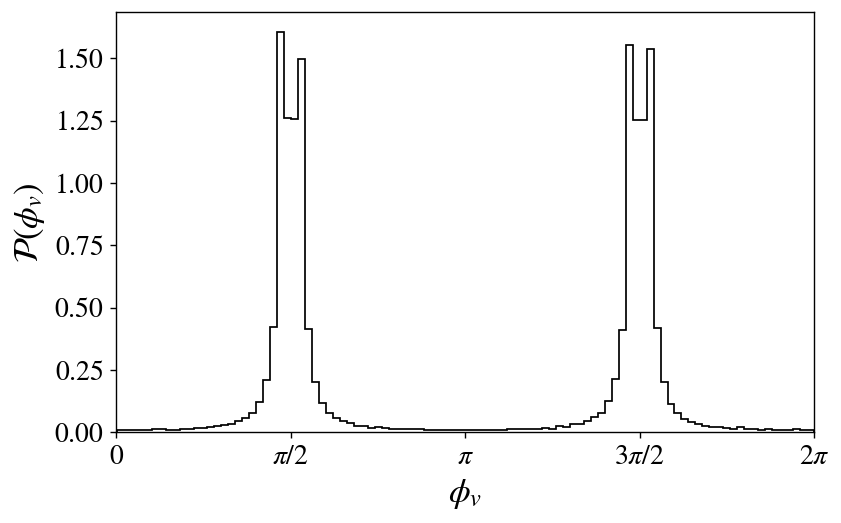}
         \caption{}
     \end{subfigure}
     \caption{$\gamma_0=1$, $M_2=0.51$, $\Delta=-0.98$, long-time behaviour shown in Fig.~\ref{fig:traj_withM_2_bic2}}     
\end{figure}

\begin{figure}[ht]
    \centering
     \begin{subfigure}[t]{.45\textwidth}
         \centering \includegraphics[height=.5\textwidth]{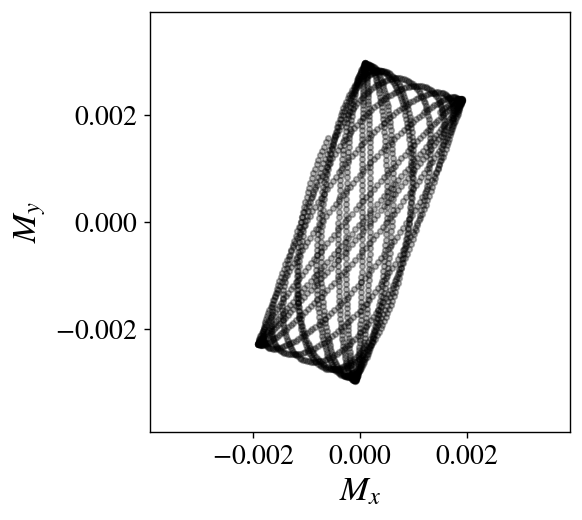}
         \caption{}
     \end{subfigure}
     \begin{subfigure}[t]{.45\textwidth}
         \centering \hspace*{-4cm}\includegraphics[height=.5\textwidth]{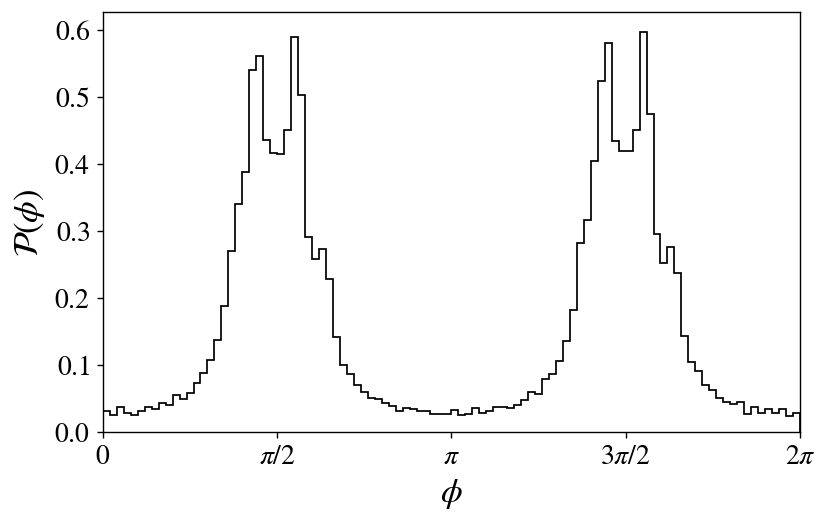}
         \caption{}
     \end{subfigure}
     \vskip\baselineskip
     \begin{subfigure}[t]{.45\textwidth}
         \centering \includegraphics[height=.5\textwidth]{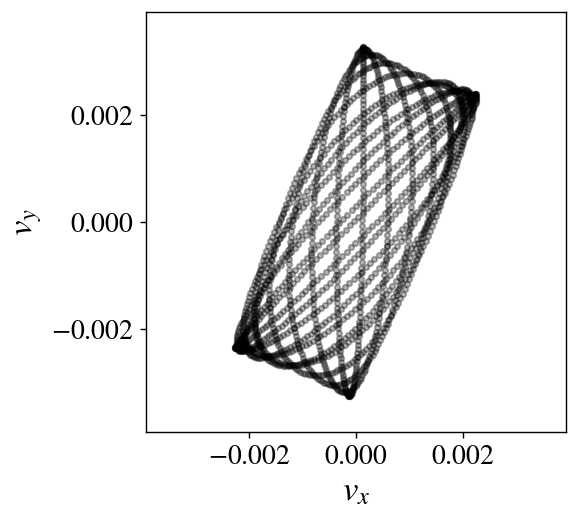}
            \caption{}
     \end{subfigure} 
     \begin{subfigure}[t]{.45\textwidth}
         \centering \hspace*{-4cm}\includegraphics[height=.5\textwidth]{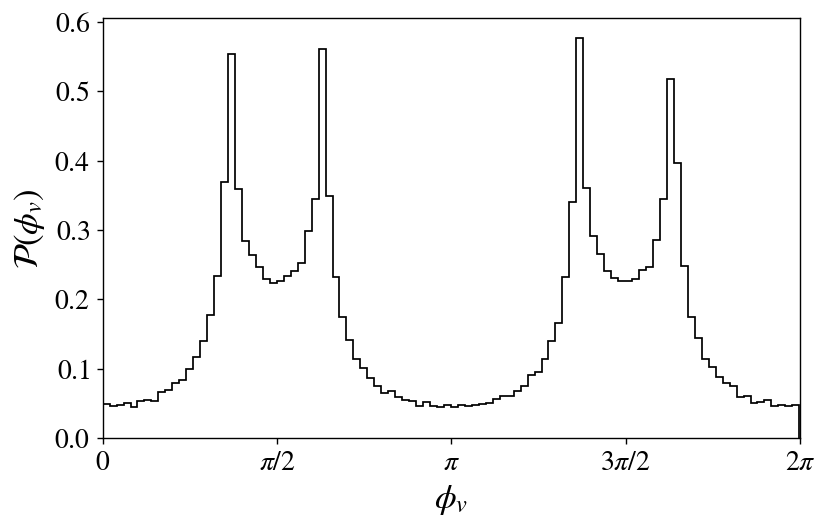}
         \caption{}
     \end{subfigure}
     \caption{$\gamma_0=0.46$, $M_2=0.06$, $\Delta=-0.94$, long-time behaviour shown in Fig.~\ref{fig:traj_withM_dis1}}
\end{figure}

\begin{figure}[ht]
    \centering
     \begin{subfigure}[t]{.45\textwidth}
         \centering \includegraphics[height=.5\textwidth]{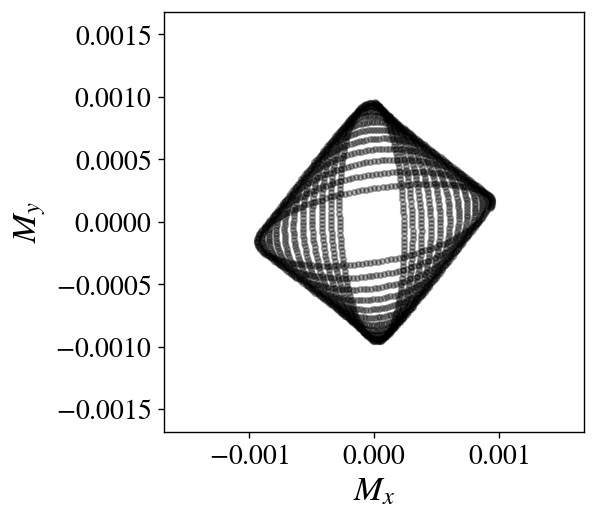}
         \caption{}
     \end{subfigure}
     \begin{subfigure}[t]{.45\textwidth}
         \centering \hspace*{-4cm}\includegraphics[height=.5\textwidth]{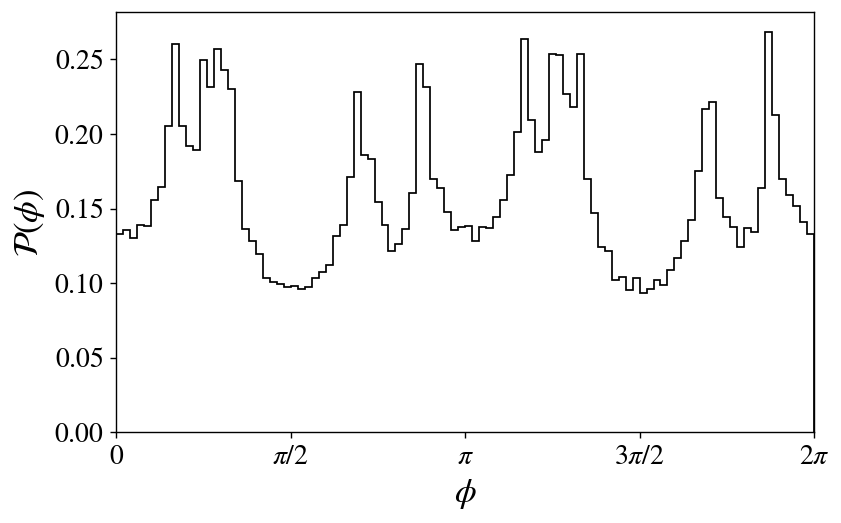}
         \caption{}
     \end{subfigure}
     \vskip\baselineskip
     \begin{subfigure}[t]{.45\textwidth}
         \centering \includegraphics[height=.5\textwidth]{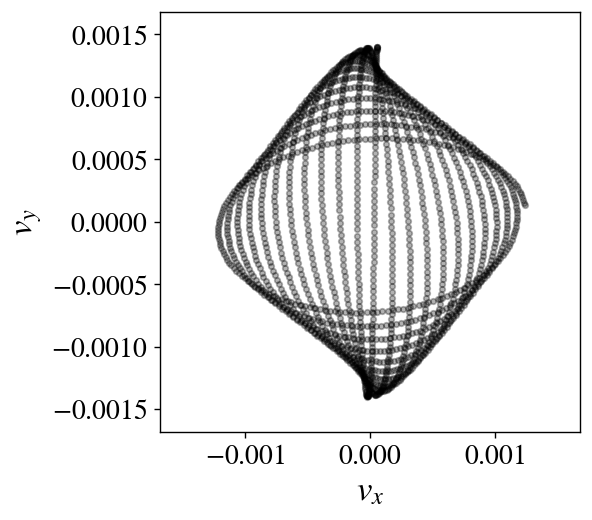}
            \caption{}
     \end{subfigure} 
     \begin{subfigure}[t]{.45\textwidth}
         \centering \hspace*{-4cm}\includegraphics[height=.5\textwidth]{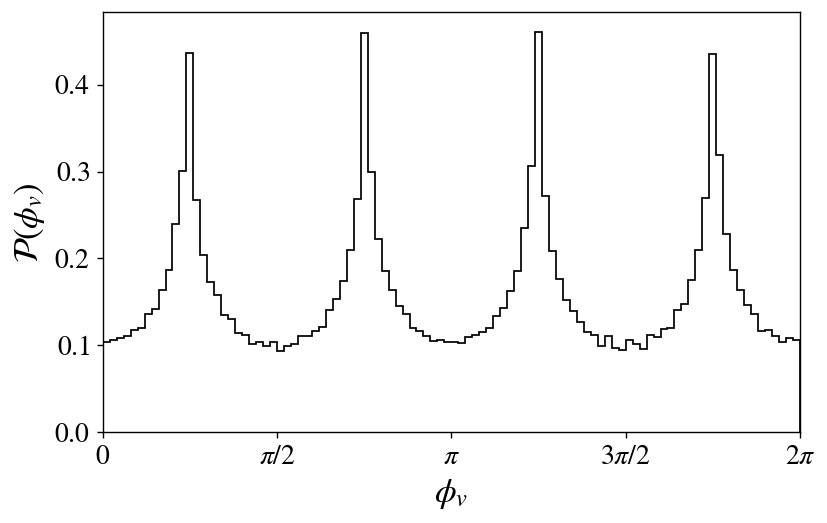}
         \caption{}
     \end{subfigure}
     \caption{$\gamma_0=0.05$, $M_2=0.03$, $\Delta=-0.11$, long-time behaviour shown in Fig.~\ref{fig:traj_withM_dis2}}
\end{figure}

\bibliography{Fast_collective_oscillations_and_clustering.bib} 

\end{document}